\documentclass[aps,prb,amsmath,amssymb,footinbib,showpacs,twocolumn]{revtex4-2}
\usepackage{amsmath}
\usepackage{amssymb}
\usepackage{amsthm}
\usepackage{setspace}
\usepackage{graphicx}
\usepackage{braket}
\usepackage{mathrsfs}
\usepackage{float}
\usepackage[colorlinks = true,linkcolor = blue,urlcolor  = blue,citecolor = blue,anchorcolor = blue]{hyperref}
\usepackage[utf8]{inputenc}
\usepackage[english]{babel}
\usepackage{bm}

\begin{document}


\title{Anomalous Transverse Response in Nodal line Semimetal Mn$_{3}$SnC}
\author{Sunil Gangwar}
\author{C. S. Yadav}
 \email{shekhar@iitmandi.ac.in}
\affiliation{School of Physical Sciences, Indian Institute of Technology Mandi, Kamand, Mandi-175075 (H.P.) India$^1$}%
\affiliation{Center for Quantum Science and Technologies, Indian Institute of Technology Mandi, Kamand, Mandi-175075 (H.P.) India$^1$}%
\date{\today}

\begin{abstract} 

The interplay of topological surface states and magnetism gives rise to the unconventional transport behaviors such as anomalous Hall effect (AHE) and anomalous Nernst effect (ANE). The antiperovskites like Mn$_3$SnC, which are nodal-line semimetal and exhibits concurrent  antiferromagnetic (AFM) - ferromagnetic (FM) ordering offer a fertile ground for anomalous transport. Here we report that the anomalous transport (AHE and ANE) in the compound is mainly dominated by the intrinsic Berry curvature effect. We establish a unified relationship between the AHE and ANE using Mott’s relation. The scattering of electrons with both antiferromagnetic and ferromagnetic magnons is manifested in the anomalous Nernst signal. The ratio of anomalous Nernst conductivity to anomalous Hall conductivity ($|\alpha^A_{xy}$/$\sigma^A_{xy}|$) is a sizable fraction of $k_{B}/e$, which points towards the stronger contribution of the Berry curvature to the ANE. 

\end{abstract}

\maketitle

\section{Introduction}
The anomalous Hall effect (AHE) and anomalous Nernst effect (ANE) have been studied in materials for both fundamental and application point of views. Unlike the ordinary Hall effect, where charge carriers get deflected by the Lorentz force, AHE features an additional contribution to the transverse voltage, arising from the spontaneous magnetization in the magnetic metals in the presence of spin-orbit coupling (SOC) \cite{nagaosa2010anomalous, shahi2022antisite,roy2020anomalous}. The ANE is the thermal counterpart of the AHE, where a transverse voltage is generated in response to a thermal gradient \cite{ghosh2021anomalous, chanda2022emergence}. These anomalous transport effects can originate from both intrinsic (due to non-zero Berry curvature) and extrinsic (such as scattering processes) factors. The band structure of the compound and thus the geometrical property of the Bloch states governs both transverse electrical and thermoelectric transport in topological materials. While the AHE probes the Berry curvature across all occupied states, the ANE is specifically related to the Berry curvature near the Fermi level \cite{xu2019large,zhang2021topological}. 
 
Recently, the antiperovskites metals such as Ca$_{3}$BiN, Ca$_{3}$PbO, Sr$_{3}$PbO, Sr$_{3}$SnO, and Mn$_{3}$ZnC have gained renewed attention due to the prediction of the topological surface states \cite{teicher2019weyl}. Generally, the topological semimetals are classified as Dirac semimetals (DSM), Weyl semimetal (WSM) and nodal line semimetal (NLSM) \cite{xu2017topological,hosen2017tunability}. The DSM and WSM feature a zero-dimensional band crossing points, while in the NLSM, the bands intersect at one-dimensional loops. The DSMs have fourfold degenerate points (Dirac nodes) near the Fermi level which are protected by time reversal symmetry (TRS) and inversion symmetry (IS) \cite{huang2024classification}. In WSM, Dirac point split into two Weyl nodes with opposite chirality due to the breaking of either TRS or IS \cite{yang2014dirac}. On the other hand, NLSM requires an additional symmetry, such as mirror symmetry, to ensure the protection of nodal lines \cite{hosen2017tunability}. A gapless NLSM often produces zero Berry-curvature, resulting in the zero value of anomalous Hall conductivity (AHC) near the Fermi level \cite{bera2023anomalous,hsieh2012topological}. However, this mirror symmetry protected gapless state can be broken in the presence of SOC, resulting in fully gapped nodal line with a pair of Weyl points \cite{bera2023anomalous}. For example, Mn$_{3}$ZnC exhibits the gapped states after the inclusion of SOC, which results in the finite values of AHC ($\sim$ 175 $\Omega^{-1} \text{cm}^{-1}$) and anomalous Nernst conductivity ($\alpha^A_{xy}$ $\sim$ 0.3 A/mK). The intrinsic Berry curvature effect and the skew scattering are reported to contribute to AHE and ANE in the different magnetic ground states of Mn$_{3}$ZnC \cite{gangwar2025berry}. Similarly, Mn$_3$GaC also exhibit drumhead like surface states, and the contribution from intrinsic Berry curvature and skew scattering \cite{gangwar2025evidences}. In addition to the Berry curvature induced ANE, several reports indicate that skew scattering can also generate a large anomalous Nernst signal, as observed in materials such as Fe$_{3}$O$_{4}$ \cite{ramos2014anomalous}, La$_{1-x}$Na$_{x}$MnO$_{3}$ \cite{ghosh2019skew}, and Co$_{1-x}$Fe$_{1+x}$CrGa \cite{chanda2022emergence}. 

The Mn$_{3}$SnC is a member of the Mn$_{3}$XC (X = Zn, Ga and Sn) family and has cubic crystal lattice (space group: Pm-3m). In these compounds, direct Mn-Mn exchange interactions give ferromagnetism whereas the Mn–C–Mn superexchange interactions lead to antiferromagnetism. The interplay between these competing interactions results in a concurrent AFM/FM state below $T_{C}$ $\sim$ 285 K in Mn$_{3}$SnC \cite{gangwar2024magneto}. Neutron diffraction studies of Mn$_3$SnC indicate that two of the three Mn atoms are arranged in a square configuration, each with an AFM moment of $\sim$ 2.3 $\mu_{B}$, while the remaining Mn atom exhibits a FM moment of $\sim$ 0.7 $\mu_{B}$ \cite{gangwar2024magneto,dias2015effect}. Owing to its isostructural relationship with Mn$_3$ZnC and Mn$_3$GaC, Mn$_3$SnC is also expected to give rise to topological surface states.

In our earlier work, we have reported structural, magnetic, magneto-transport and thermoelectric properties of Mn$_{3}$SnC \cite{gangwar2024magneto}. Here we explore the AHE and ANE in the light of non-trivial band structure of the compound. The estimated temperature dependence of the $\alpha^A_{xy}/\sigma^A_{xy}$ suggest a value which is a sizable fraction of $k_{B}/e$ at high temperature. At low temperature (T = 2 K), low values of AHC and ANE are observed, approximately  0.85 $\Omega^{-1} \text{cm}^{-1}$ and 8.1 $\times$ $10^{-4}$ A/mK, respectively. We also calculate the anomalous Hall and anomalous Nernst angles, both of which are observed to be very low. 

\section{Results and discussion}

\subsection{Anomalous Hall Effect}

 \begin{figure*}
\includegraphics[width= 17 cm, height = 9 cm]{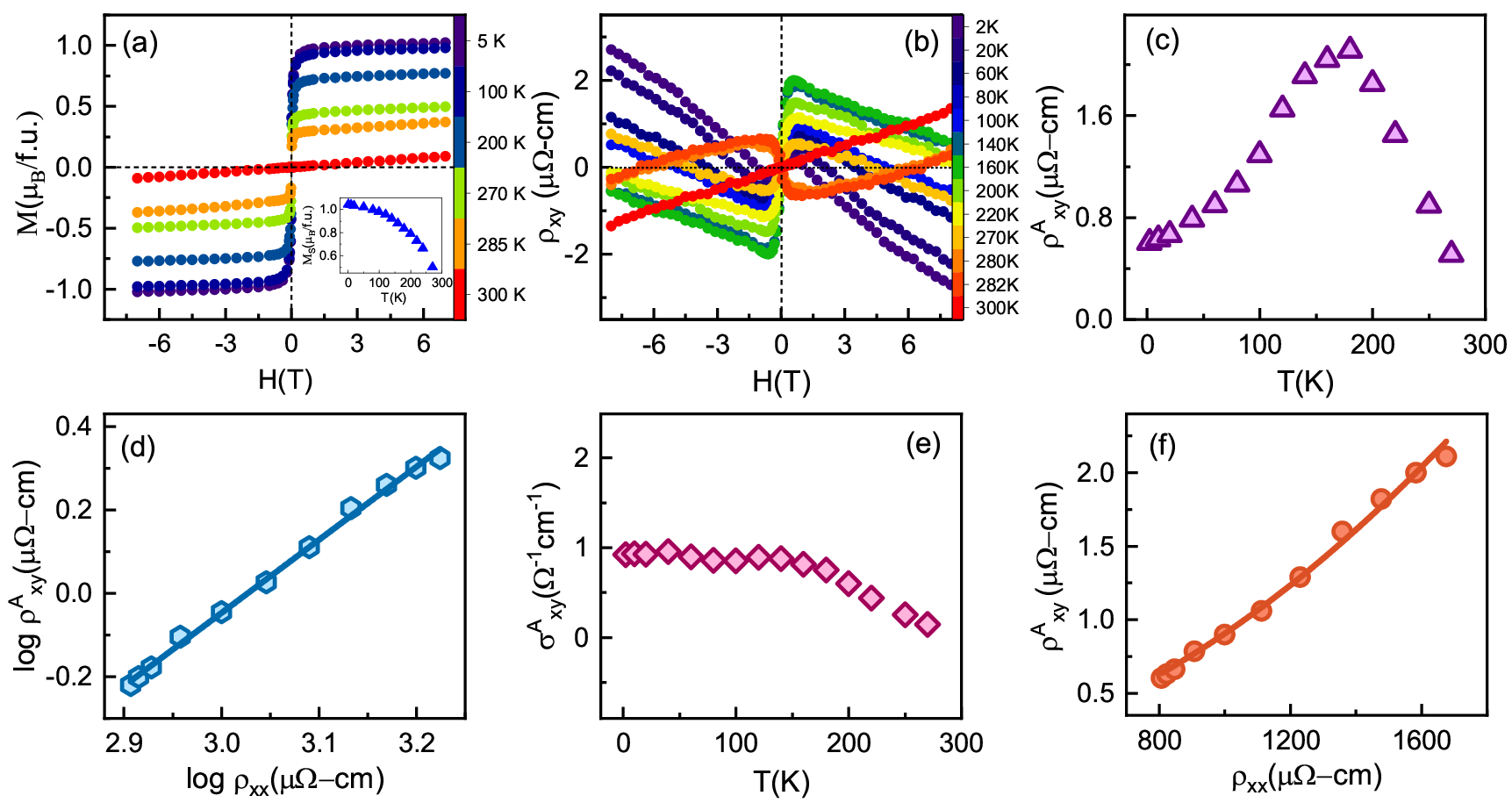}
\caption{(a) Magnetic field dependent magnetization M at different temperatures. (b) Magnetic field dependent Hall resistivity $\rho_{xy}$ at different temperatures. (c) Temperature dependent anomalous Hall resistivity $\rho^A_{xy}$. (d) The plot of log $\rho^A_{xy}$  and log $\rho_{xx}$ data, solid black line indicating the fit using the relation $\rho^A_{xy} \propto \rho^\gamma_{xx}$. (e) Temperature dependent anomalous Hall conductivity $\sigma^A_{xy}$. (f) The plot between $\rho^A_{xy}$ and $\rho_{xx}$, solid line representing the fit using Eq.(3).}
\label{fig:Figure1}
\end{figure*}

We performed magnetotransport measurements in the temperature range of 2 – 300 K up to an 8 T field to explore the AHE. The Hall resistivity ($\rho_{xy}$) data were antisymmetrized using the expression $\rho_{xy} = [\rho_{xy}(H) - \rho_{xy}(-H)]/2$ to eliminate the longitudinal component, the resulting values were plotted as a function of the applied magnetic field in Fig. 1(b). Below $\sim$ 1 T field, the $\rho_{xy}(H)$ curves increase rapidly with the field, similar to the isothermal magnetization data, indicating the AHE in the system (see Fig. 1(a)). In FM systems, the Hall resistivity includes an additional contribution from spontaneous magnetization, in addition to the ordinary Hall resistivity, and can be written as \cite{shukla2022band,shahi2022antisite,bera2023anomalous,chatterjee2023nodal,nagaosa2010anomalous}
\begin{equation}
 \rho_{xy} (H) = \rho^0_{xy} + \rho^A_{xy} = R_{0}H + R_{S}M_{S}   
\end{equation}
where $\rho^A_{xy}$, $R_{0}$, $R_{S}$ and $M_{S}$ represent the anomalous Hall resistivity, ordinary Hall coefficient, anomalous Hall coefficient and saturated magnetization respectively.

The AHE is mainly understood though intrinsic Berry curvature mechanism or the extrinsic mechanisms. The intrinsic mechanism is linked to the interband scattering of carriers under SOC, leading to the generation of an anomalous velocity in addition to the group velocity, which results in the AHC in the system. Later, it was reinterpreted in terms of Berry curvature, which acts as a fictitious magnetic field and influences the motion of charge carriers. The extrinsic mechanism includes the skew scattering and side jump of the charge carriers. In skew scattering, spin-polarized charge carriers scatter from impurities or defects in the presence of SOC, which causes a deviation from their original paths. Such asymmetric scattering is responsible for the AHC \cite{roy2020anomalous,shahi2022antisite,bera2023anomalous,chatterjee2023nodal,nagaosa2010anomalous}. The Side-jump is a quantum effect, in which the trajectory of charge carriers is shifted by a distance $\Delta$x from their original path, leading to a distortion of the wave function, which contributes to the AHC \cite{berger1970side}.

The values of $R_{0}$ and $\rho^A_{xy}$ are extracted from the linear fit of $\rho_{xy} (H)$ curves at the high field region. The slope and the y-axis intercept of the linear fit represent $R_{0}$ and $\rho^A_{xy}$, respectively \cite{shukla2022band,shahi2022antisite,bera2023anomalous,chatterjee2023nodal}. The $R_{0}$ and carrier concentrations ($n$) were discussed in our previous report \cite{gangwar2024magneto}. Figure 1(c) presents the temperature dependence of $\rho^A_{xy}$. As the temperature increases from 2 K to 180 K, $\rho^A_{xy}(T)$ increases and reaches a maximum value of about 2.1 $\mu\Omega$ cm. With further increase in temperature, $\rho^A_{xy}$ decreases, owing to the reduction of the saturated magnetic moment near 200 K.

To elucidate the origin of the observed AHE in Mn$_{3}$SnC, we examine the exponent ($\gamma$) using the scaling relation $\rho^A_{xy} \propto \rho^\gamma_{xx}$, the value of $\gamma$ = 1 and 2 indicating the skew scattering and intrinsic Berry curvature or side contributions, respectively \cite{shahi2022antisite,bera2023anomalous,chatterjee2023nodal,shukla2022atomic}. We believe that above $\sim$ 200 K, magnetic transition starts to influence the behavior of $\rho^A_{xy}$, and it start to decrease rapidly. So, we restrict our analysis in the temperature range of $T$ = 2 - 180 K for a good comparison \cite{bera2023anomalous}. We plot $\rho^A_{xy}$ and $\rho_{xx}$ on a double logarithmic scale, and a linear fit was used to determine the value of $\gamma$. The fitted value of $\gamma$ is found to be 1.75 $\pm$ 0.02, as shown in Fig. 1(d). This suggests the contributions from both the intrinsic Berry curvature and the skew scattering mechanisms to AHE. Further, we calculate the AHC ($\sigma^A_{xy}$), from the observed values of $\rho^A_{xy}$ and $\rho_{xx}$ as \cite{shahi2022antisite,bera2023anomalous,chatterjee2023nodal,shukla2022atomic}
\begin{equation}
   \sigma^A_{xy} = \frac{\rho^A_{xy}}{(\rho^A_{xy})^2 + (\rho_{xx})^2}
\end{equation}
 The $\sigma^A_{xy} (T)$ decreases as the temperature increases (Fig 1(e)), reaching a peak value of $\sim$ 0.85 $\Omega^{-1}\text{cm}^{-1}$ at 2 K. In this compound, FM and AFM phases coexist and compete with each other down to low temperature. Since the intrinsic AHC originates from the Berry curvature integrated over the Brillouin zone, where each k-point contributes with a sign, the net Berry curvature depends sensitively on the underlying symmetries. A finite Berry curvature generally requires the breaking of either time-reversal or inversion symmetry. In this system, the competing FM and AFM interactions tend to partially restore symmetry macroscopically. As a result, Berry curvature contributions from different sublattices and electronic bands largely cancel each other, leading to the small values of AHC.
 The $\sigma^A_{xy} (T)$ does not show an appreciable change in the entire temperature range, indicating a scattering-independent induced AHE. This suggests that the dominant contribution comes from the intrinsic Berry curvature. The overall variation of the AHC is very small, amounting to only $\sim$ 0.7 $\Omega^{-1}$ $cm^{-1}$ over the investigated temperature range. The AHC remains nearly temperature independent up to $\sim$ 200 K, while only a weak temperature dependence is observed in the range 200 $\le$ $T$ $\le$ 270\,\text{K}. This weak variation may originate from additional scattering contributions associated with defects and/or phonon-magnon scattering processes.
 To separate out the intrinsic and skew scattering contributions, we use the conventional scaling model to fit the experimental data  \cite{bera2023anomalous,chatterjee2023nodal}
\begin{equation}
 \rho^A_{xy} = a^{skew}\rho_{xx} + \sigma^{int}_{xy}\rho_{xx}^2  
 \end{equation}
where $a^{skew}$ and $\sigma^{int}_{xy}$ represent the contribution of skew scattering and intrinsic to the AHC, respectively. We plot $\rho^A_{xy}$ as a function of $\rho_{xx}$ in Fig. 1(f), fitted values of $a^{skew}$ and $\sigma^{int}_{xy}$ are found to be 0.0001 and $\sim$ 0.71 $\Omega^{-1}\text{cm}^{-1}$, respectively. Thus, intrinsic Berry curvature contribution have $\sim$ 80 \% contribution of the total AHC. By using the fitted values of $a^{skew}$ and $\sigma^{int}_{xy}$, we have shown separate plots for the skew scattering and intrinsic Berry curvature contributions to $\rho^A_{xy}$ in Fig. 2(a). It is evident that the intrinsic Berry curvature contribution dominates over the skew scattering contribution across the entire temperature range. 

Furthermore, the anomalous Hall angle ($\theta_{AH}$) and anomalous Hall coefficient ($S_{H}$) are calculated, which exhibit relative strength of the AHE in a compound \cite{chatterjee2023nodal}. The $\theta_{AH}$ is defined as the ratio of AHC and longitudinal conductivity ($\theta_{AH} = \sigma^A_{xy}/\sigma_{xx}$), which represents the fraction of the longitudinal current that is converted into the anomalous Hall current, due to non-zero Berry curvature. The $S_{H}$ shows the relation between AHC and saturation magnetization $M_{S}$, defined as $S_{H} = \rho^A_{xy}/M_{S}$. In Berry-curvature induced AHE, $S_H$ exhibits a temperature-independent behavior, while in skew-scattering-induced AHE it shows a strong temperature dependence. \cite{bera2023anomalous,chatterjee2023nodal}. In Fig. 2(b), we plot $\theta_{AH}$ and $S_{H}$ as a function of temperature. The $\theta_{AH} (T)$ increases with rising temperature, reaching a peak value of $\sim$ 0.1 \% near 180 K, and then decreases, following a trend similar to $\rho^A_{xy}( T)$. Whereas, $S_{H} (T)$ shows nearly temperature independent behavior, which further reflect the Berry curvature induced AHE.  

\subsection{Anomalous Nernst Effect}
 
In FM systems, the total Nernst signal is sum of the ordinary Nernst signal ($S^0_{xy}$) and anomalous Nernst signal ($S^A_{xy}$),  as \cite{li2023enhanced,uchida2021transverse} 
\begin{equation}
S_{xy} (H) = S^0_{xy} + S^A_{xy} = Q_{0}H + Q_{S}M_{S}
\end{equation}
 where $Q_{0}$ and $Q_{S}$ are the normal Nernst coefficient and anomalous Nernst coefficient. The ordinary Nernst effect refers to the generation of a transverse voltage in the presence of longitudinal temperature gradient and perpendicular applied magnetic field. The ANE is a magnetization dependent phenomenon that arises from SOC and non-trivial band structure of the material. The anti-symmetrized $S_{xy}(H)$ data is plotted in Fig. 3(a) as a function of applied field at different temperatures. The $S_{xy}$ evolves with applied magnetic field similarly to $\rho_{xy}(H)$ displaying a sharp increase at low field followed by a gradual decrease at higher field. In Fig. 3(b), we plot $S^A_{xy}$ as a function of temperature, obtained by extrapolating of high-field $S_{xy}(H)$ curve to zero field on the y-axis \cite{uchida2021transverse}. The $S^A_{xy}(T)$ shows a strong temperature dependence with two broad peaks at $\sim$ $60\text{K}$ and $\sim$ $160\text{ K}$. The $T$ $\sim 60\text{ K}$ peak is associated with either phonon-drag or magnon-drag, as discussed in previous work, while the $T$ $\sim 160\text{ K}$ peak can be linked to the reduction in the saturated magnetic moments, similar to what is observed in the $\rho^A_{xy}(T)$ \cite{gangwar2024magneto}. The maximum value of $S^A_{xy}(T)$ $\sim$ 0.012 $\mu$\text{V/K} at $\sim$ 160 K, is comparable to CoFeCrGa (0.018 $\mu$\text{V/K}) \cite{chanda2022emergence}, SrRuO$_{3}$ films (0.03 $\mu$\text{V/K}) \cite{kan2019strain}, CrRuMnGe (0.04 $\mu$\text{V/K}) \cite{chanda2024large} and CoFeVSb (0.039 $\mu$\text{V/K}) \cite{chanda2023intrinsic}. 

 \begin{figure}
	\begin{center}
		\includegraphics[width=7.0cm]{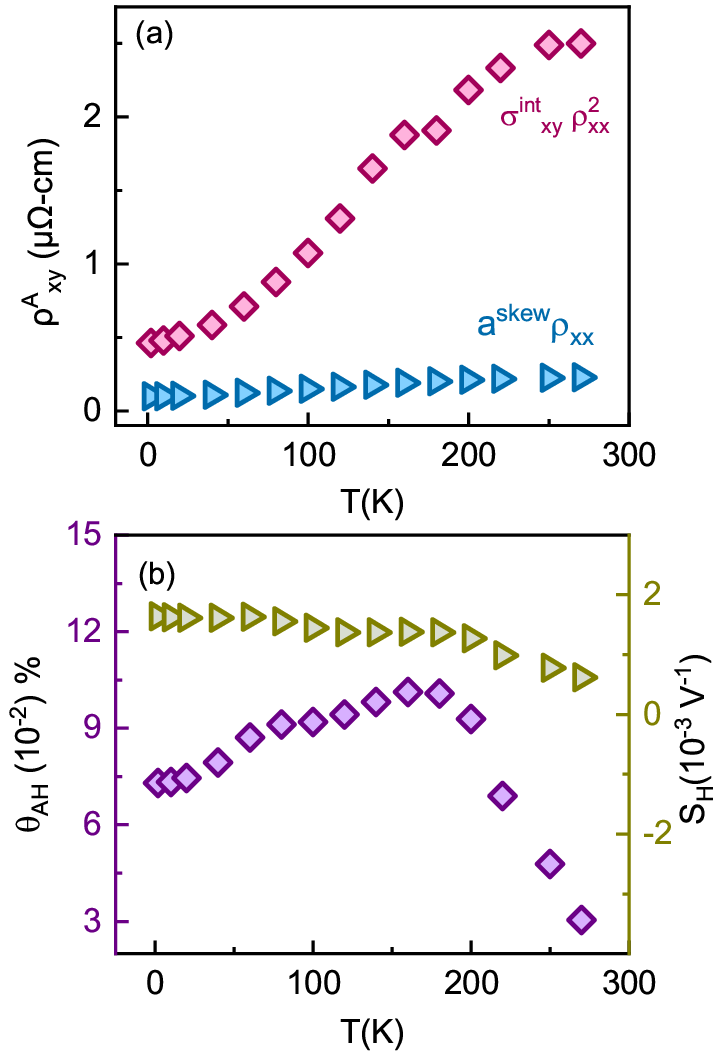}
		\caption{\label{Fig2}(a) Extrinsic ($a^{skew}\rho_{xx}$) and intrinsic ($\sigma^{int}_{xy}\rho^2_{xx}$) contribution in $\rho^A_{xy}$ as a function of temperature. (b) Temperature dependent anomalous Hall angle $\theta_{AH}$ and anomalous scaling coefficient $S_{H}$.}
	\end{center}
\end{figure}

  \begin{figure*}
\includegraphics[width= 17 cm, height = 9 cm]{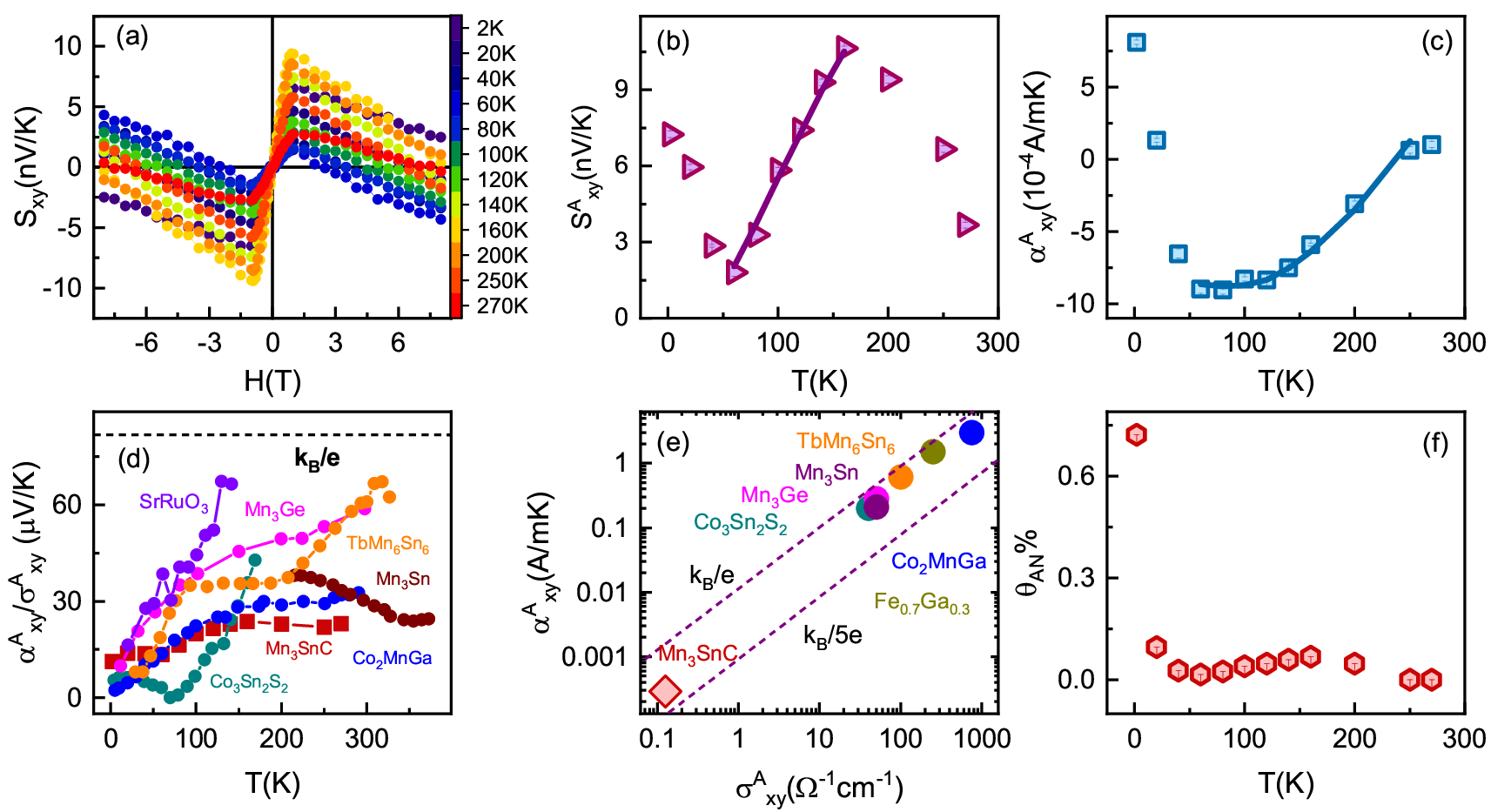}
\caption{(a) Magnetic field dependent Nernst signal $S_{xy}$ at different temperatures. (b) Temperature dependent anomalous Nernst signal $S^A_{xy}$, solid line representing the fit using Eq. (6). (c) Temperature dependent anomalous Nernst conductivity $\alpha^A_{xy}$, solid line representing the fit using Eq. (7). (d) The ratio $|\alpha^A_{xy}/\sigma^A_{xy}|$ as a function of temperature in different magnets including Mn$_{3}$Sn \cite{ikhlas2017large}, Mn$_{3}$Ge \cite{xu2020finite}, SrRuO$_{3}$ \cite{miyasato2007crossover}, Co$_{2}$MnGa \cite{xu2020anomalous}, Co$_{3}$Sn$_{2}$S$_{2}$ \cite{ding2019intrinsic}, TbMn$_{6}$S$_{6}$ \cite{wei2025large} and Mn$_{3}$SnC. (e) Plot between $\alpha^A_{xy}$ and $\sigma^A_{xy}$ for Mn$_{3}$SnC and other reported materials at room temperature (Mn$_{3}$SnC and Co$_{3}$Sn$_{2}$S$_{2}$ are taken at 270 K and 170 K, respectively because of the magnetic order. (f) Anomalous Nernst angle $\theta_{AN}$ as a function of temperature.}
\label{fig:Figure2}
\end{figure*}

The ANE can also originate from both intrinsic Berry curvature and extrinsic mechanisms (skew scattering and side jump). In general, $S^A_{xy}$ can be described through its correlation with the longitudinal electrical conductivity $\sigma_{xx}$, $S_{xx}$, $\sigma^A_{xy}$ and transverse thermoelectric conductivity (Nernst conductivity) $\alpha^A_{xy}$, as \cite{chanda2023intrinsic,li2023enhanced,chanda2024large}
\begin{equation}
 \alpha^A_{xy} = \sigma_{xx}S^A_{xy} + \sigma^A_{xy}S_{xx}   
\end{equation}
 
The values of $\sigma_{xx}$ and $S_{xx}$ were calculated in our previous study \cite{gangwar2024magneto}. In Fig. 3(c), we plot $\alpha^A_{xy}$ as a function of temperature, which exhibits a maximum value of about $\sim$ 0.01 A/mK at 2 K. The observed behavior of $\alpha^A_{xy} (T)$ in Mn$_{3}$SnC is similar to that reported for Mn$_{3}$Sn \cite{li2017anomalous}. Using the Mott's relation $S_{xx}$ and $\alpha^A_{xy}$ can be expressed in term of energy derivative of $\sigma_{xx}$ and $\sigma^A_{xx}$ at the Fermi level as $S_{xx} = \frac{\pi^2 k^2_B T}{3e \sigma_{xx}} \left( \frac{\partial \sigma_{xx}}{\partial E} \right)_{E=E_F}$ and $\alpha^A_{xy} = \frac{\pi^2 k^2_B T}{3e} \left( \frac{\partial \sigma^A_{xy}}{\partial E} \right)_{E=E_F}$, where $E_{F}$ is the Fermi energy. Considering the power law for AHE, $\rho^A_{xy} = \lambda \rho^{\gamma}_{xx}$, where $\lambda$ represents SOC constant. In terms of these parameters, $S^A_{xy}$ can be written as \cite{chanda2024large,ramos2014anomalous,chanda2023intrinsic}
\begin{equation}
  S^A_{xy} = \rho^\gamma_{xx} \left[ \frac{\pi^2 k^2_{B} T}{3e} \left( \frac{\partial \lambda}{\partial E} \right)_{E=E_F} - (\gamma - 1)\lambda S_{xx} \right]
\end{equation}

Here, the value of $\gamma = 1$ suggests that the ANE is governed by the skew scattering mechanism, while $\gamma = 2$ indicates dominance of intrinsic Berry curvature or side-jump mechanisms \cite{chanda2024large,ramos2014anomalous,chanda2023intrinsic}. Using $\rho_{xx}(T)$ and $S_{xx}(T)$ data, $S^A_{xy}(T)$ is fitted in the temperature range of 60 - 160 K by employing Eq. (6), with $\left( \frac{\partial \lambda}{\partial E} \right)_{E=E_F}$, $\lambda$, and $\gamma$ as fitting parameters. Due to the presence of two broad peaks at $\sim$ 60 K and $\sim$ 160 K, the fitting was restricted to the intermediate temperature range. The fitted curve, represented by the solid blue line in Fig. 3(b), yields a value of $\gamma = 1.6 \pm 0.023$. To further elucidate the origin of the observed ANE, we examine the temperature dependence of $\alpha^A_{xy}(T)$ in term of $\rho_{xx}$ and $S_{xx}$. According to the Mott's relation $\alpha^A_{xy}$ can be expressed as 
\begin{equation}
  \alpha^A_{xy} = \rho^{(\gamma-1)}_{xx} \left[ \frac{\pi^2 k^2_{B} T}{3e} \left( \frac{\partial \lambda}{\partial E} \right)_{E=E_F} - (\gamma - 2)\lambda S_{xx} \right]   
\end{equation}
We performed a fit of $\alpha^A_{xy}(T)$ over the temperature range of 60 K to 220 K, and the best fit was obtained for $\gamma = 1.71 \pm 0.039$, which closely matches the value obtained from the $S^A_{xy} (T)$ fit. The fitted values of $\gamma$ obtained from $S^A_{xx}(T)$ and $\alpha^A_{xy}$ are comparable to those observed for the AHE, indicating that the ANE is predominantly driven by intrinsic Berry-curvature effects. The ratio [$(\frac{\partial \lambda}{\partial E})_{E=E_F}/\lambda$] obtained from $S^A_{xy} (T)$ and $\alpha^A_{xy}(T)$ fits are $1.33 \times 10^{19}$ and $1.33 \times 10^{19}$, respectively, which is comparable to other reported compounds \cite{ramos2014anomalous,chanda2024large}. 

Furthermore, the $\alpha^A_{xy}$ and $\sigma^A_{xy}$ can be expressed in term of Berry curvature $\Omega_{B}$ as \cite{xu2020anomalous,li2023enhanced,asaba2021colossal,yang2020giant}
\begin{equation}
    \sigma^A_{xy} = \frac{e^2}{ \hbar(2\pi)^3} \int_{\text{BZ}} d^3 k \, f(k) \, \Omega_B(k) \approx \frac{e^2}{\hbar} \frac{1}{a} <\frac{\Omega_B}{\lambda^2_{F}}>
\end{equation}
\begin{equation}
 \alpha^A_{xy} = \frac{e k_{B}}{\hbar(2\pi)^3} \int_{\text{BZ}} d^3 k \, s(k) \, \Omega_B(k) \approx \frac{e k_B}{\hbar} \frac{1}{a} <\frac{\Omega_B}{\Lambda^2}>
\end{equation}
where $k_{B}$, $e$, $\hbar$, $a$, $f(k)$ and $s(k) = -f(k)ln([(k)]-[1-f(k)]ln[1-f(k)]$, $\lambda_{F}$ and $\Lambda$ represent Boltzmann's constant, electric charge, Planck's constant, lattice parameter, Fermi-Dirac distribution function, entropy density of the electron gas, Fermi wave length and de Broglie thermal wavelength, respectively. We can simplify the expression by dividing Eq.(9) by Eq.(8), leading to the result $|\frac{\alpha^A_{xy}}{\sigma^A_{xy}}| \approx \frac{k_B}{e} < \frac{\lambda_F^2}{\Lambda^2} >$ and $\lambda_{F}$ can be written in term of Fermi radius; $\lambda_{F} = 2\pi/k_{F}$ \cite{li2023enhanced,xu2020anomalous,chanda2023intrinsic}.  The value of $|\alpha^A_{xy}/\sigma^A_{xy}|$ lies between 0.2 $k_{B}/e$ and 0.9 $k_{B}/e$ for all known topological materials \cite{xu2020anomalous,asaba2021colossal}. The $|\alpha^A_{xy}/\sigma^A_{xy}|$ ratio approaches fraction of $k_{B}/e$ in Berry curvature induced ANE, as reported for Co$_{2}$MnGa, SmMnB2$_{2}$ and UCo$_{0.8}$Ru$_{0.2}$Al  \cite{chanda2023intrinsic,xu2020anomalous,asaba2021colossal}.  In Fig. 3(d), we plot the ratio $|\alpha^A_{xy}/\sigma^A_{xy}|$ as a function of temperature with other reported topological materials. The temperature variation of $|\alpha^A_{xy}/\sigma^A_{xy}|$ for Mn$_{3}$SnC is similar to that reported for Co$_{3}$Sn$_{2}$S$_{2}$ \cite{yang2020giant}, increases with increasing the temperature. The maximum value of $|\alpha^A_{xy}/\sigma^A_{xy}|$ for Mn$_{3}$SnC is found to be $\sim$ 20 $\mu \text{V}/\text{K}$ (0.23 $k_{B}/e$) which is comparable to the other materials. Further, we have shows the values of $\alpha^A_{xy}$ as a function of $\sigma^A_{xy}$ at the room temperature in Fig. 3(e); except for the Co$_{3}$Sn$_{2}$S$_{2}$ (170 K) \cite{ding2019intrinsic} and Mn$_{3}$SnC (270 K) due to the magnetic ordering. The value of $|\alpha^A_{xy}/\sigma^A_{xy}|$ for all the compounds lie between $k_{B}/e$ to $k_{B}/5e$ as shown by dashed violet and pink line respectively for high temperature data. These observations suggest that the mechanism underlying the ANE in Mn$_3$SnC is consistent with other magnetic topological materials, confirming its Berry curvature driven nature.

Similar to $\theta_{AH}$, we calculate anomalous Nernst angle ($\theta_{AN}$), defined as the ratio of $S^A_{xy}$ and the $S_{xx}$ \cite{chuang2017enhancement,hu2018anomalous}. It decreases with decreasing temperature  down to $\sim$ 50 K. Above this temperature, it exhibits a hump-like feature at $\sim$ 160 K, consistent with $S^A_{xy}$ data, and reaches a maximum value of $\sim$ 0.72 \%. The sharp increase below $\sim$50 K can be attributed to the enhanced phonon-drag effect, which becomes increasingly prominent as the temperature decreases within the expected temperature regime.

\section{conclusion}
In summary, we present an investigation of the AHE and ANE in the magnetic topological nodal line semimetal Mn$_{3}$SnC. Our findings indicate that both the AHE and ANE in Mn$_3$SnC originate from the intrinsic Berry-curvature mechanism. The anomalous Nernst signal reflects magnon or phonon drag contributions around 50 K. We demonstrate the common connection between AHE and ANE through Mott's relation. Our results reveal the correlation between $\alpha^A_{xy}$ and $\sigma^A_{xy}$, and the ratio $\alpha^A_{xy}/\sigma^A_{xy}$ is found to be a sizable fraction of $k_{B}/e$, indicating Berry curvature effect in the material. We have found a very low value of AHC $\sim$ 0.85 $\Omega^{-1} \text{cm}^{-1}$ at 2 K with an intrinsic contribution of $\sim$ 0.71 $\Omega^{-1} \text{cm}^{-1}$. The maximum values of $S^A_{xy}$ and $\alpha^A_{xy}$ are $\sim$ 1.2 n\text{V} $\text{K}^{-1}$ and $\sim$ 1 $\times$$10^{-3}$A/mK respectively. Moreover, we have observed very low values 0.1 \% and 0.8 \% of $\theta_{AH}$ and $\theta_{AN}$ respectively. Due to the coexistence of FM and AFM order, the FM component enhances the Berry curvature, whereas the AFM component suppresses it, resulting in a reduced net Berry curvature compared to a purely FM system. Consequently, the low values of $S^A_{xy}$  and $\alpha^A_{xy}$ are attributed to the weak contribution from the Berry curvature.

\section{Acknowledgment}
We acknowledge Advanced Material Research Center (AMRC), IIT Mandi for the experimental facilities. SG and CSY acknowledge IIT Mandi and India for the HTRA fellowship. This research received no external funding.

\bibliography{Mn3SnC}

\end{document}